\documentclass[aps,pra,superscriptaddress,floatfix,showpacs,showkeys,10pt,twocolumn]{revtex4}
\usepackage[dvips]{graphicx}
\usepackage{amsmath,amsfonts,amssymb,graphics,graphicx,epsfig,color,times,bbm}

\begin{document}

\title{Implementation of quantum algorithms with resonant interactions}
\author{Zhuo-Liang Cao}
\email{zlcao@ahu.edu.cn}
\author{Ping Dong}
\email{pingdong@ahu.edu.cn}

\affiliation{Key Laboratory of Opto-electronic Information
Acquisition and Manipulation, Ministry of Education, School of
Physics and Material Science, Anhui University, Hefei, 230039, P R
China} 

\begin{abstract}
We propose a scheme for implementing quantum algorithms with
resonant interactions. Our scheme only requires resonant
interactions between two atoms and a cavity mode, which is simple
and feasible. Moreover, the implementation would be an important
step towards the fabrication of quantum computers in cavity QED
system.
\end{abstract}

 \maketitle

Construction of quantum computer is an enormously appealing task
because of quantum computational potential to perform superfast
quantum algorithms. Two classes of quantum logarithms illustrate the
great theoretical promise of quantum computers. One is based on
Shor's Fourier transformation including quantum factoring
\cite{Shor}, Deutsh-Jozsa logarithm \cite{D} and so on, which are
all exponential speedup compared with the classical algorithms. The
other is based on Grover's quantum search \cite{Grover}, which is
quadratic speedup compared with the classical ones. The Grover
search algorithm is very important because many techniques based on
the search algorithm are universally used in our lives. The Grover
search algorithm is efficient to look for one item in an unsorted
database of size $N\equiv2^{n}$ \cite{Grover,H}. Classically, in
order to achieve the task, one needs $O(N)$ queries.  However, one
needs $O(\sqrt{N})$ queries by the Grover search algorithm.
Furthermore, the efficiency of the algorithm has been manipulated
experimentally in few-qubit cases via Nuclear Magnetic Resonance
(NMR) \cite{1,2} and optics \cite{3,4}.

Here, we first review the general Grover search algorithm. The
circuit diagram of the Grover search algorithm with $n$ data qubits
and one auxiliary working qubit, which can be used to search one
item from $2^{n}$ items, is shown in Fig. {\ref {1}}. The process
can be concluded as the following seven steps:
\begin{enumerate}
    \item Prepare the $n+1$ qubits, which are in $|0\rangle^{\otimes n}|1\rangle_{n+1}$;
    \item Perform the $n+1$ Hadamard transformations on the $n+1$ qubits ;
    \item Apply the oracle. The auxiliary working qubit can be omitted after the step;
    \item Perform the $n$ Hadamard transformations on the $n$ data qubits;
    \item Apply a phase shift to the data qubits except$|0\rangle^{\otimes n}$,
          which can be described by the unitary operator $2|0\rangle^{\otimes n}\langle0|-I$
          where $I$ is the identity operation on the data qubits;
    \item Perform the $n$ Hadamard transformations on the $n$ data qubits again;
    \item Repeat steps $3\rightarrow 6$ with a finite number of times, then measure the
    $n$ data qubits. \\
\end{enumerate}
The number of repetitions \cite{M} for obtaining a finite item is
$R=CI (\frac{arccos\sqrt{1/N}}{2arccos \sqrt{N-1/N}})$ , which is
bounded above by $\pi \sqrt{N}/4$.

\begin{figure}[tbp]
\includegraphics[scale=0.30,angle=90]{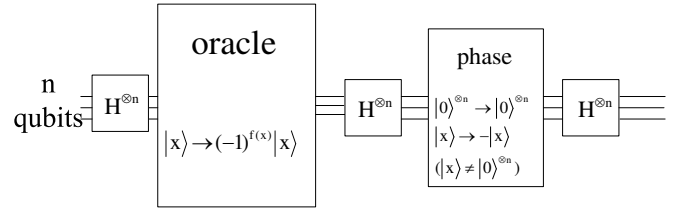}
\caption{The circuit diagram for the Grover search algorithm.
$H^{\otimes n}$ denotes $n$ Hadamard transformations on the $n$ data
qubits. An auxiliary working has been omitted. If $x$ is the one to
be searched, $f(x)=1$, otherwise, $f(x)=0$. }\label{fig1}
\end{figure}

On the other hand, in the realm of atom, cavity QED techniques,
where atoms interact with a quantized electromagnetic field, have
been proved to be a promising candidate for realizing the quantum
processors. Recently, many schemes of quantum algorithms have been
proposed based on cavity QED techniques. For example, Rauschenbeutel
\emph{et al} \cite{Rauschenbeutel}  have been realized a two-qubit
phase gate experimentally with resonant interaction of a two-level
atom with a cavity mode and Zheng \cite{Zheng} has realized a
two-qubit controlled-phase gate with resonant interaction of two
three-level atoms with a cavity mode. The Deutsh-Jozsa (D-J)
logarithm \cite{Zheng1} and the Grover search algorithm
\cite{Yamaguchi,Deng} have been realized in cavity QED, and so on.

In this paper, we first mainly propose a simple scheme for
implementing the Grover search algorithm in cavity QED. Comparing
the Refs \cite{Yamaguchi,Deng}, they are both based on non-resonant
interactions, our scheme is based on single resonant interactions
between atoms and cavity and does not use the cavity mode as the
data bus. Thus the current scheme is very simple and the interaction
time is very short, which is important in view of decoherence. More
importantly, we strictly investigate the case of atomic spontaneous
emission and cavity decay during the interactions. Then we avoid the
effect by constructing appropriate unitary transformations.
Therefore our proposal is more approach to real case and can succeed
with higher fidelity (over $0.99$). Here, we only discuss the case
of two data qubits, where we can search a finite item from four
items. The circuit diagram of the Grover search with two data qubits
and one auxiliary working qubit is shown in Fig. {\ref {2}}.
Three-level atoms are used in this paper and the relevant level
structure is shown in Fig. \ref {3}. The third level $|i\rangle$ is
not affected during the atom-cavity resonant interaction. Thus we
consider the case that two atoms interact with the single mode
cavity mode, in the interaction picture, the Hamiltonian of the
atom-cavity interaction can be expressed as (assuming $\hbar=1$)
\cite{Zheng}
\begin{figure}[tbp]
\includegraphics[scale=0.30,angle=90]{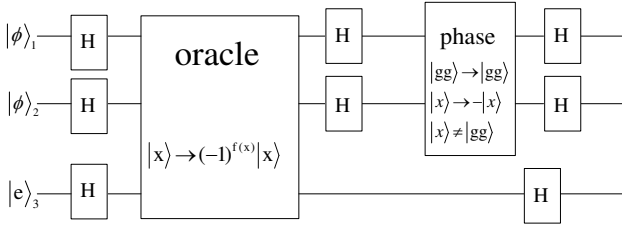}
\caption{The circuit diagram for the two-data-qubit Grover search
algorithm. $H$ denotes Hadamard transformation. $|\phi\rangle_{1}$
and $|\phi\rangle_{2}$ are two data qubits, $|e\rangle_{3}$ is an
auxiliary working qubit.} \label{fig2}
\end{figure}

\begin{equation}
\label{1}
H=g_{1}(a^{\dagger}S_{1}^{-}+aS_{1}^{+})+g_{2}(a^{\dagger}S_{2}^{-}+aS_{2}^{+}),
\end{equation}
where $g_{1}$ and $g_{2}$ are the coupling strength of the atoms 1,
2 with the cavity, respectively. $s^{+}=|e\rangle\langle g|$,
$s^{-}=|g\rangle\langle e|$ and $|g\rangle$ is the ground state of
the atoms, $|e\rangle$ is the excited state of the atoms.
$a^{\dagger}$, $a$ are the creation and annihilation operators of
the cavity mode. Assume that the cavity mode is initially prepared
in the vacuum state $|0\rangle_{c}$. In order to implement the
two-data-qubit Grover search algorithm, firstly, we prepare atoms 1,
2 and 3 in the state

\begin{figure}[tbp]
\includegraphics[scale=0.15,angle=0]{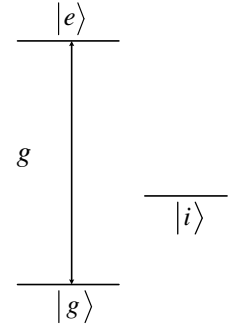}
\caption{The level structure of the atoms. $|g\rangle$ is the ground
state, $|e\rangle$ is the excited state. The cavity mode is
resonantly coupled to the $|e\rangle\leftrightarrow |g\rangle$
transition. The third level $|i\rangle$ is not affected by the
interaction.} \label{fig3}
\end{figure}

\begin{equation}
\label{2} |\phi\rangle_{123}=|gge\rangle_{123}.
\end{equation}
and send atoms 1 and 3 through a classical field and choose
appropriately phase and amplitude, respectively
\begin{subequations}
\label{3}
\begin{equation}
 |g\rangle_{1} \rightarrow
\frac{1}{\sqrt{2}}(|g\rangle_{1}+|e\rangle_{1}),
\end{equation}
\begin{equation}
|e\rangle_{3} \rightarrow
\frac{1}{\sqrt{2}}(|g\rangle_{3}-|e\rangle_{3}),
\end{equation}
\end{subequations}
Then we send atom 2 through two classical fields and choose
appropriately phases and amplitudes
\begin{equation}
\label{4} |g\rangle_{2} \rightarrow
\frac{1}{\sqrt{2}}(|g\rangle_{2}+|e\rangle_{2})\rightarrow
\frac{1}{\sqrt{2}}(|g\rangle_{2}+|i\rangle_{2}).
\end{equation}
So the total state of the atoms 1 , 2 and 3 becomes
\begin{equation}
\label{5} |\phi\rangle_{123}=\frac{1}{2\sqrt{2}}(|gg\rangle_{12}+
|gi\rangle_{12}+|eg\rangle_{12}+|ei\rangle_{2})(|g\rangle_{3}-|e\rangle_{3}).
\end{equation}

Obviously, we have known the four items ( $|gg\rangle_{12}$ ,
$|gi\rangle_{12}$ , $|eg\rangle_{12}$ , $|ei\rangle_{2}$ ) are
stored in the data qubits before applying the Oracle. Without loss
of generality, we search the state $|eg\rangle_{12}$ from the four
states. However, for the two-data-qubit Grover search algorithm, the
oracle has effect on the states to be searched. The auxiliary
working qubit can be discarded at this point.

Secondly, we send atoms 1 and 2 through the vacuum cavity, the
evolutions are governed by the Hamiltonian of Eq. (\ref{1}),
\begin{subequations}
\label{6}
\begin{eqnarray}
|eg\rangle_{12}|0\rangle_{c}\rightarrow\frac{g_{1}}{E}\{\frac{1}{E}(g_{1}cos(Et)+\frac{g_{2}^{2}}{g_{1}})|eg\rangle_{12}|0\rangle_{c}\nonumber\\
+\frac{1}{E}g_{2}[cos(Et)-1]|ge\rangle_{12}|0\rangle_{c}-i
sin(Et)|gg\rangle_{12}|1\rangle_{c}\},
\end{eqnarray}
\begin{equation}
|ei\rangle_{12}|0\rangle_{c}\rightarrow[cos(g_{1}t)|e\rangle_{1}|0\rangle_{c}-i
sin(g_{1}t)|g\rangle_{1}|1\rangle_{c}]|i\rangle_{2},
\end{equation}
\begin{equation}
|gg\rangle_{12}|0\rangle_{c}\rightarrow|gg\rangle_{12}|0\rangle_{c},
\end{equation}
\begin{equation}
|gi\rangle_{12}|0\rangle_{c}\rightarrow|gi\rangle_{12}|0\rangle_{c},
\end{equation}
\end{subequations}
where $E=\sqrt{g_{1}^{2}+g_{2}^{2}}$. If we choose
\begin{equation}
\label{7} t=\frac{\pi}{g_{1}},  g_{2}=\sqrt{3}g_{1},
\end{equation}
which can be achieved by choosing coupling strengths and
interaction time appropriately. Thus, we have
\begin{subequations}
\label{8}
\begin{equation}
|eg\rangle_{12}|0\rangle_{c}\rightarrow
|eg\rangle_{12}|0\rangle_{c},
\end{equation}
\begin{equation}
|ei\rangle_{12}|0\rangle_{c}\rightarrow
-|ei\rangle_{12}|0\rangle_{c},
\end{equation}
\begin{equation}
|gg\rangle_{12}|0\rangle_{c}\rightarrow
|gg\rangle_{12}|0\rangle_{c},
\end{equation}
\begin{equation}
|gi\rangle_{12}|0\rangle_{c}\rightarrow
|gi\rangle_{12}|0\rangle_{c}.
\end{equation}
\end{subequations}
Then send atom 2 through two classical fields tuned to the
transition
\begin{equation}
\label{9} |i\rangle_{2}\rightarrow |e\rangle_{2},
|g\rangle_{2}\longleftrightarrow |e\rangle_{2}.
\end{equation}
These lead the state of atoms 1 and 2 to
\begin{equation}
\label{10} |\phi\rangle_{12}=\frac{1}{2}(|gg\rangle_{12}+
|ge\rangle_{12}-|eg\rangle_{12}+|ee\rangle_{12}).
\end{equation}

Thirdly, we send atoms 1 and 2 through a classical field,
respectively. Choosing appropriately phase and amplitude, let
\begin{subequations}
\label{11}
\begin{equation}
 |g\rangle_{i} \rightarrow
\frac{1}{\sqrt{2}}(|g\rangle_{i}+|e\rangle_{i}),(i=1,2)
\end{equation}
\begin{equation}
|e\rangle_{i} \rightarrow
\frac{1}{\sqrt{2}}(|g\rangle_{i}-|e\rangle_{i}).(i=1,2)
\end{equation}
\end{subequations}
Then we perform a single-qubit operation on atom 2 again
\begin{equation}
\label{12} |e\rangle_{2}\rightarrow |i\rangle_{2}.
\end{equation}
Thus Eq. (\ref{10}) becomes
\begin{equation}
\label{13} |\phi\rangle_{12}=\frac{1}{2}(|gg\rangle_{12}+
|eg\rangle_{12}-|gi\rangle_{12}+|ei\rangle_{12}).
\end{equation}

In order to achieve the next step (phase transformation), we can
perform single-qubit operations and controlled-phase
transformations on the two atoms as in Eq. (\ref{8}), which can
lead Eq. (\ref{13}) to
\begin{equation}
\label{14} |\phi\rangle_{12}=\frac{1}{2}(|gg\rangle_{12}+
|ge\rangle_{12}-|eg\rangle_{12}-|ee\rangle_{12}).
\end{equation}

Finally, we perform single-qubit operations on atoms 1 and 2 as in
Eq. (\ref{11}). Thus we obtain the state of atoms 1 and 2
\begin{equation} \label{15}
|\phi\rangle_{12}= |eg\rangle_{12}.
\end{equation}
We can measure the state of atoms 1 and 2 by detectors. Obviously,
the state of atoms 1 and 2 is the result that we want to search. If
we want to search other states ($|gg\rangle_{12}$, $|gi\rangle_{12}$
or $|ei\rangle_{12}$), the main process is the same as above (Eq.
(\ref{8}) and Eq. (\ref{11})), except for some single-qubit
operations.

But in the real processing of resonant interactions, the cavity
decay and the atomic spontaneous emission are unavoidable. Taking
them into consideration, if we choose appropriate parameters
$g_{2}=\sqrt{3}g_{1}$, $t=\frac{\pi}{g_{1}}$ and
$\kappa=\tau=0.1g_{1}$, where $\kappa$ is the cavity decay rate and
$\tau$ is the atomic spontaneous emission rate, the evolution
\cite{Zheng} of system is similar to Eq. (\ref{8}). While
$10^{-\pi/20}$ is added to the $|eg\rangle_{12}$ and
$-|ei\rangle_{12}$ compared with the ideal case. The state of Eq.
(\ref{14}) becomes

\begin{eqnarray}
\label{16}
|\phi\rangle_{12}&=&\frac{1}{\sqrt{1+2\times10^{-\frac{\pi}{10}}+10^{-\frac{\pi}{5}}}}
(|gg\rangle_{12}-
10^{-\frac{\pi}{20}}|ge\rangle_{12}\nonumber\\&&+10^{-\frac{\pi}{20}}|eg\rangle_{12}-10^{-\frac{\pi}{10}}|ee\rangle_{12}).
\end{eqnarray}

Then we perform the single-qubit operations
\begin{subequations}
\label{17}
\begin{equation}
 |g\rangle_{1} \rightarrow
\frac{1}{\sqrt{1+10^{-\pi/10}}}(10^{-\pi/20}|g\rangle_{1}+|e\rangle_{1}),
\end{equation}
\begin{equation}
|e\rangle_{1} \rightarrow
\frac{1}{\sqrt{1+10^{-\pi/10}}}(|g\rangle_{1}-10^{-\pi/20}|e\rangle_{1}),
\end{equation}
\end{subequations}
and
\begin{subequations}
\label{18}
\begin{equation}
 |g\rangle_{2} \rightarrow
\frac{1}{\sqrt{1+10^{-\pi/10}}}(|g\rangle_{2}+10^{-\pi/20}|e\rangle_{2}),
\end{equation}
\begin{equation}
|e\rangle_{2} \rightarrow
\frac{1}{\sqrt{1+10^{-\pi/10}}}(10^{-\pi/20}|g\rangle_{2}-|e\rangle_{2}),
\end{equation}
\end{subequations}
on atoms 1 and 2, respectively. These lead the state of atoms 1 and
2 to $|eg\rangle_{12}$, \emph{i.e.,} we can search the state
perfectly ( the successful possibility and fidelity are both equal
to 1.0 ).

Out of question, the D-J algorithm can be also implemented with
resonant interactions. The D-J algorithm can distinguish the
function $f(x)$ between constant and balanced \cite{D}. The values
of the function $f(x)$ are either 0 or 1 for all possible inputs.
The values of balance function are equal to 1 for half of all the
possible inputs, and 0 for the other half. The constant is always 1
or 0 for all inputs. Classically, if we want to unambiguously
distinguish between constant and balanced function on $2^{n}$
inputs, we will need $2^{n}/2+1$ queries to achieve the task. While
for the D-J algorithm, we will need only one query. Here we discuss
the two-qubit D-J algorithm. The state of query and auxiliary
working qubit is prepared in $(|0\rangle_{i} +
|1\rangle_{i})(|0\rangle_{a}-|1\rangle_{a})/2$. Then mapping a
unitary transformation $U_{f}$ on the system, which will lead the
initially state to
$[(-1)^{f(0)}|0\rangle_{i}+(-1)^{f(1)}|1\rangle_{i}]
(|0\rangle_{a}-|1\rangle_{a})/2$. There are four possible
transformations to the $U_{f}$: (1) for $U_{f1}$, $f(0)=f(1)=0$; (2)
for $U_{f2}$, $f(0)=f(1)=1$; (3) for $U_{f3}$, $f(0)=0$ and $f(1) =
1$; (4) for $U_{f4}$, $f(0)=1$ and $f(1)=0$. After a Hadamard
transformation on the query qubit, the state of query qubit becomes
$|f(0)\oplus f(1)\rangle$. If the function $f(x)$ is constant, the
state of query qubit becomes $|0 \rangle_{i} $. Otherwise it becomes
$|1\rangle_{i}$.

Obviously implementation of the unitary transformation $U_{f}$ is
the key. We prepare two atoms in
\begin {equation}
|\varphi\rangle_{12}=(|g\rangle_{1} +
|e\rangle_{1})(|g\rangle_{2}-|e\rangle_{2})/2.
\end{equation}

In the case of $U_{f1}$, we take no operation on the two atoms.

In the case of $U_{f2}$, we perform a single-qubit rotation on atom
2,
\begin {equation}
\label{18}|g\rangle_{2}\longleftrightarrow|e\rangle_{2}.
\end{equation}

In the case of $U_{f3}$, we first perform a single-qubit rotation on
atom 2
\begin {equation}
\label{19}|e\rangle_{2}\longleftrightarrow|i\rangle_{2}.
\end{equation}
Secondly we send atoms 1 and 2 through a vacuum cavity. we can
obtain the evolution of Eq. (\ref{8}), which is governed by the
Hamiltonian of Eq. (\ref{1}). Thirdly we perform single-qubit
rotation on atom 2 of Eq. (\ref{19}) and another single-qubit
rotation of Eq. (\ref{18}). Then we send atoms 1 and 2 through the
vacuum cavity again as Eq. (\ref{8}). Finally, we perform
single-qubit rotation on atom 2 of Eq. (\ref{19}).

In the case of $U_{f4}$, we can achieve the task by the process of
the case of $U_{f3}$ and a single-qubit operation on atom 2.

Now, we have completed the unitary $U_{f}$. Thus the two-qubit D-J
algorithm will be implemented simply. Moreover, the scheme can be
generalized to multi-qubit case.

For the real processing (with cavity decay and the atomic
spontaneous emission), we can also achieve the task with successful
possibility and fidelity are both equal to 1.0 by choosing
appropriate sing-qubit operations as the implementation of Grover
search algorithm. This is not only a useful character in
experimental manipulation but also important for constructing real
quantum computer.

Discussion on the feasibility of the current scheme is necessary.
The scheme requires two atoms in a vacuum cavity have different
coupling strengths with the cavity mode. The coupling depends on the
atomic positions: $g=\Omega e^{-r^{2}/\omega^{2}}$, where $\Omega$
is the coupling strength at the cavity center, $\omega$ is the waist
of the cavity mode, and $r$ is the distance between the atom and the
cavity center \cite{p}. The condition $g_{2}=\sqrt{3}g_{1}$ in our
scheme can be satisfied by locating one atom at the center of the
cavity and locating the other one at the position $r=\omega
ln^{1/2}\sqrt{3}$. According to the recent experiments with Cs atoms
trapped in an optical cavity\cite{v}, the condition can be obtained.
For the resonant cavity, in order to implement quantum algorithms
successfully, the relationship between the interaction time and the
excited atom lifetime should be taken into consideration. The
interaction time should be much shorter than that of atom radiation.
Hence, atom with a sufficiently long excited lifetime should be
chosen. For Rydberg atoms with principal quantum numbers $50$ and
$51$, the radiative time is $T_{1}\simeq3\times10^{-2}$s . From the
analysis in Ref \cite{biao}, the interaction time is on the order
$T\simeq2\times10^{-4}$s , which is much shorter than the atomic
radiative time. So the condition can be satisfied by choosing
Rydberg atoms. Furthermore our scheme requires that two atoms be
simultaneously sent through a cavity, otherwise there will be an
error. Assume that during the interaction between the two atoms and
the cavity, one atom enters the cavity $0.01t$ sooner than another
atom, with $t$ being the time of each atom staying in the cavity. We
can obtain the fidelity $F\simeq 0.999$ for Eq. (\ref{10}) and the
total fidelity is about 0.998 for the two data qubits grover
algorithm. Obviously in this case the operation is only slightly
affected in current schemes.

Next, one needs to reach the Lamb-Dicke regime in order to implement
the quantum algorithms. For the state of Eq. (\ref{5}), in the
Lamb-Dicke regime, the infidelity caused by the spatial extension of
the atomic wave function is about $\Delta\simeq (ka)^{2}\pi$, where
$k$ is the wave vector of the cavity mode and $a$ is the spread of
the atomic wave function. Setting $\Delta\simeq0.01$, so we have
$a\simeq0.01\lambda$, where $\lambda$ is the wavelength of the
cavity mode. If the atom trajectories cross the cavity with the
deviation of less $0.1$ degree from its pre-determined direction, we
can ensure the fidelity is about $0.999$ for Eq. (\ref{10}) and the
unitary transformation $U_{f}$ . While in order to maintain
$g_{2}=\sqrt{3}g_{1}$ in the process of atomic motion in the cavity,
we can choose the parameter of cavity $z\leq0.5z_{0}$, where
$z_{0}=\frac{\pi\omega^{2}}{\lambda}$ and $2z$ is the length of the
cavity. We can obtain the error is only about $10^{-3}$. Therefore
our scheme is feasible with the current cavity QED technology.

In conclusion, we have proposed the scheme for implementing the
quantum algorithms in cavity QED. Our scheme only requires resonant
interactions between two atoms and a cavity mode. The interaction
time is very short, which is very important in view of decoherence.
Meanwhile, even if we take the cavity decay and atomic spontaneous
emission into consideration, we can still achieve the task
perfectly. Moreover, the implementation of the algorithms would be
an important step to scale more complex quantum algorithms and our
scheme would be very important for constructing real quantum
computer.

\textbf{Acknowledgments}
This work is supported by the National
Natural Science Foundation of China under Grant No. 60678022, the
Doctoral Fund of Ministry of Education of China under Grant No.
20060357008, the Key Program of the Education Department of Anhui
Province under Grant Nos: 2006KJ070A, 2006KJ057B and the Talent
Foundation of Anhui University.

\end{document}